\documentclass[aps,preprint]{revtex4}%
\usepackage{amsfonts}
\usepackage{amsmath}
\usepackage{amssymb}
\usepackage{graphicx}%
\begin{document}
\title{Configurations of a new atomic interferometer for gravitational wave detection}
\author{Biao Tang$^{a,b}$}
\author{Baocheng Zhang$^{a}$}
\email{zhangbc@wipm.ac.cn}
\author{Lin Zhou$^{a}$}
\author{Jin Wang$^{a}$}
\email{wangjin@wipm.ac.cn}
\author{Mingsheng Zhan$^{a}$}
\email{mszhan@wipm.ac.cn}
\affiliation{$^{a}$State Key Laboratory of Magnetic Resonances and Atomic and Molecular
Physics, Wuhan Institute of Physics and Mathematics, The Chinese Academy of
Sciences, Wuhan 430071, China}
\affiliation{$^{b}$University of Chinese Academy of Sciences, Beijing 100049, China}

\begin{abstract}
Recently, the configuration using atomic interferometers (AIs) had
been suggested for the detection of gravitational waves. A new AI
with some additional laser pulses for implementing large momentum
transfer was also put forward, in order to improve the influence of
shot noise and laser frequency noise. In the paper, we use the
sensitivity function to analyze all possible configurations of the
new AI and to distinguish how many momenta are transferred in a
specific configuration. With the analysis for the new configuration,
we explore the detection scheme of gravitational wave further, in
particular, for the amelioration of the laser frequency noise. We
find that the amelioration is definite in such scheme, but novelly,
in some cases the frequency noise can be canceled completely by
using a proper data processing method.

\end{abstract}
\maketitle


\section{Introduction}

Since gravitational waves were predicted from the theory of General
Relativity, a direct detection of gravitational waves had become an exciting
frontier of experimental physics \cite{gvlb13}. The initial detection scheme
was related to the resonant bar detectors \cite{jw69} which would be
mechanically disturbed by passing gravitational waves, and then some other
detectors which had one or more arms with its arm length influenced by passing
gravitational waves were put forward. For the latter, the VIRGO \cite{aaa10}
and LIGO \cite{aaa09} are the most sensitive detectors of this class as an
kind of interferometer modulated by the change of the apparent distance
between mirrors by passing gravitational waves. Many efforts had also been
made to improve the frequency range and the sensitivity of the possible probed
gravitational waves, e.g. Advanced LIGO \cite{gmh10}, and a large space-based
laser interferometer gravitational wave detector, LISA \cite{ep96}. Although
there is no any definite and direct evidence to show the existence of
gravitational waves up to now, these efforts gave an upper limit on the
stochastic gravitational-wave background of cosmological origin \cite{lv09}
which was considered as a realistic observable tied directly to the
quantization of gravity in a recent paper \cite{kw13}.

Another direction of efforts to detect gravitational waves is
related to the use of atomic coherence. The first scheme along this
direction is MIGO (Matter-wave Interferometric Gravitational-wave
Observatory) \cite{cs04,sc04} in which the atom wave is used with
the same role as the photons in LIGO. In particular, it was shown
\cite{rbp06} recently that the scheme based on MIGO would no better
than LIGO, which made the attention focused on the MIGO reduced to
some extent. However, the appearance of MIGO stimulated the thoughts
on using atomic interferometers to detect gravitational waves. An
interesting design along the line is AGIS (Atomic Gravitational Wave
Interferometric Sensor) \cite{dgr08} which works with a similar
mechanism to LIGO but replacing the macroscopic mirrors with freely
falling atoms. The use of freely falling atoms or atomic
interferometers as local inertial sensors reduced the requirement of
elaborate seismic isolation which limits the sensitivity of LIGO at
Hz-band or lower frequencies, so the AGIS have an advantage over
LIGO by detecting gravitational waves between 0.01 and 100 Hz.
However, in AGIS, the Raman process in the atomic interferometers
still induces an uncontrollable noise through the laser phase
fluctuations which is also a dominant noise background for the
detection of gravitational waves using LIGO. The reason of existence
of laser frequency noises in AGIS \cite{dghk08} is the use of two
laser beams in the interaction with the atoms, in which the pulses
from the control laser are common to both interferometers and the
phase contributions from this laser would be canceled in the final
differential phase shift, but the noise in the phase of the passive
laser cannot be canceled completely. This leads to a recent
suggestion \cite{yt10} for the gravitational wave detection using
the similar structure to AGIS but replacing the local inertial
sensors with single-laser atom interferometers, which overcomes the
laser frequency noise existed in the gravitational wave detection
using LIGO but still inherits the advantages of AGIS to the
suppression of vibration noise.

However, the measurement of interferometry is also limited by the
shot noise which is related to the number of particles attending the
interference process. Thus following the optimistic assumption
\cite{dgr08} about the number of atom when operating an atomic
interferometer, the shot noise is still larger in an atomic
interferometer than in an optical interferometer. In particular, the
typical number of photon impinging on a photo detector is easily
increased than the number of atoms in an atomic interferometer. Thus
besides developing a new technology to increase the number of atoms,
one has to use the large momentum transfer (LMT) \cite{dgr08,mclh08}
to compensate for the low number of atoms. Based on this background,
a new method for gravitational wave detection with atomic sensors
was recently suggested \cite{ghkr13}, which not only overcame the
laser frequency noise using single-laser atomic interferometers, but
also realized the LMT by adding some laser pulses between the basic
beam splitter and mirror pulses. However, in Ref. \cite{ghkr13}, the
authors gave only a prototypical illumination for the mechanism
realizing LMT, and didn't make the detailed analysis for the
possible configurations of the new interferometer. This forms the
main purpose of this paper, that is to analyze the configurations of
the new interferometer and the difference of the momentum transfer
for different configurations. On the other hand, we will also
attempt to analyze the cancellation of laser frequency noises with a
direct and explicit mathematical expression.

In order to analyze the structure of new atomic interferometers, we introduce
the sensitivity function \cite{ccl05} to distinguish the different momentum
transfer for different structures and thus the complex calculation for the
final total phase difference using the standard method of non-relativistic
quantum mechanics can be avoided. The second section of the paper will give a
brief introduction of sensitivity function, in particular for its application
to sense the signal about the gravitational field. In the third section, we
analyze the possible configurations for the new interferometers and find some
interesting phenomena for some configurations. In the fourth section, we
present the results of laser frequency noise cancellation with an analysis of
the total laser phase difference. In particular, we find the method of the
time-delay interferometry (TDI) used in Ref. \cite{yt10} to cancel the laser
frequency noise is not applicable to the scheme suggested in Ref.
\cite{ghkr13}. Finally, we summarize our conclusions in the fifth section.

\section{Sensitivity function}

In this section we will revisit the sensitivity function, without loss of
generality, in the time-domain atomic interferometer proposed firstly by
Kasevich and Chu \cite{kc91}. The interferometer consists of beam
splitter-mirror-beam splitter ($\frac{\pi}{2}-\pi-\frac{\pi}{2}$) optical
pulse sequence, and its sensitivity is limited to a large extent by the phase
noise derived from the lasers as well as the residual vibrations. Similar to
the Ref. \cite{ccl05}, our investigation for the sensitivity function here is
under the assumption of short laser pulses with the description of pure plane waves.

In the interferometer we considered here, the output of the results is
presented by the population change of atomic number, e.g. the probability of
finding the atom remained in the ground state when leaving the interferometer
is $P=\frac{1+\cos\left(  \Delta\Phi\right)  }{2}$ where $\Delta\Phi
=\Delta\Phi_{s}+\Delta\Phi_{n}$ is the total phase difference
\cite{dghk08,kc92,sc94,pcc97,zcz13} between the two paths of the
interferometer, which is also the basis of experimental observation. In a
local gravitational measurement around the Earth, the leading order of the
signal can be calculated as $\Delta\Phi_{s}=kgT^{2}$ where $k$ is the
effective laser-field wavevector, $T$ is the interrogation time between two
sequent laser pulses, and $g$ is the local gravitational acceleration. And
$\Delta\Phi_{n}=\phi_{1}-2\phi_{2}+\phi_{3}$ is the interferometric phase from
the interaction between three laser pulses and atoms, and it is usually locked
to the value $\frac{\pi}{2}$ such that the transition probability is $\frac
{1}{2}$ which ensures the highest sensitivity to any interferometric phase
fluctuations. From the perspective of the configuration, the signal can also
be regarded as derived from a certain kind of interference, e.g. its influence
is included in the interferometric phase $\Delta\Phi_{n}$. Thus it is not hard
to understand why the noises usually constrain the sensitivity of the
interferometer. In the following we will consider $\Delta\Phi=\Delta\Phi_{n}$
and $\Delta\Phi_{s}$ enters into the total phase difference as the influence
of gravitational field on the interferometric phase. In particular, we will
consider the same interferometer but with a single laser pulse to interact
with the atom, and thus the Raman process is replaced with a single photon
transition and the redundant effects related to the ac Stark shifts would
disappear \cite{wyc94,pcc01}. In this case, the signal, expressed by the
leading order phase shift in a local gravitational field, is proportional to
the atomic energy level difference, which will be seen later.

As suggested firstly by Dick \cite{gjd87} and then investigated in detail by
Cheinet, et al \cite{ccl05}, the sensitivity of a time-domain atomic
interferometer can be characterized by the sensitivity function which
quantifies the influence of a relative laser phase shift $\delta\phi$
occurring at a time $t$ during the interferometer sequence onto the transition
probability $\delta P\left(  \delta\phi,t\right)  $; it is then defined in
Ref. \cite{gjd87} as%
\begin{equation}
g\left(  t\right)  =2\ast\underset{\delta\phi\rightarrow0}{\lim}\frac{\delta
P\left(  \delta\phi,t\right)  }{\delta\phi}. \label{sf}%
\end{equation}
If the time origin is chosen at the middle of the second Raman pulse, the
sensitivity function $g\left(  t\right)  $ is an odd function. For the three
pulses $\frac{\pi}{2}-\pi-\frac{\pi}{2}$ with durations respectively
$\tau-2\tau-\tau$, we choose the initial time $t_{i}=-T$ and the final time
$t_{f}=T$ to get the expression of the sensitivity function as%
\begin{equation}
g(t)=\left\{
\begin{array}
[c]{c}%
\sin(\Omega(T+t))\ \ \ ,-T\leqslant t<-T+\tau\\
\ 1\ \ \ \ \ \ \ \ \ \ \ \ \ ,-T\ +\tau\leqslant t<-\tau\\
-\sin\Omega t\ \ \ \ \ \ \ \ ,-\tau\leqslant t<\tau\ \ \ \ \ \ \\
-1\ \ \ \ \ \ \ \ \ \ \ \ ,\tau\leqslant t<T-\tau\ \ \ \\
-\sin\Omega(T-t)\ \ \ \ ,T-\tau\leqslant t\leqslant T\ \ \
\end{array}
\right.  \label{sf1}%
\end{equation}
where $\Omega$ is the effective Rabi frequency and $g(t)=0$ for $\left\vert
t\right\vert >T$ due to the phase jump occurs outside the interferometer. As
seen in Fig.1, the sensitivity function $g(t)$ is indeed an odd function, so
in the analysis of the next section we will present a sensitivity function
with the part only for $t<0$. Usually the Fourier transform of the sensitivity
function is required, since the noises related closely to the analysis of the
sensitivity of an interferometer can be expressed in terms of a power spectral
density which is expanded with the frequency. Thus the transfer function which
is introduced in the appendix1 is usually used in the analysis of the
sensitivity, but in the present paper we focus on the analysis of the
structure of the interferometer, so the sensitivity function is enough. We
will also use the transfer function when we discuss the advantages of some
kinds of interferometers, e.g. the transfer of the influence of the noises.

\bigskip\begin{figure}[ptb]
\centering
\includegraphics[width=3.25in]{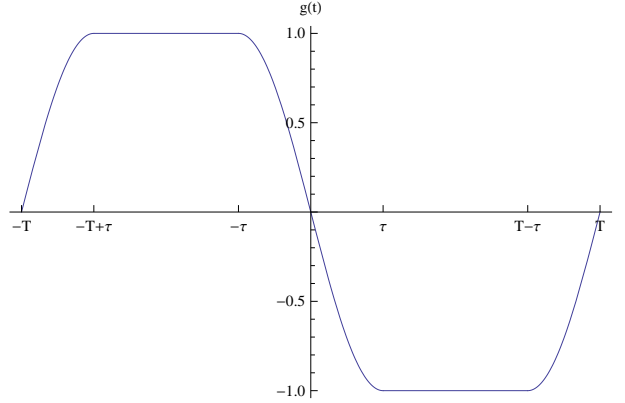}\caption{the sensitivity function g(t) }%
\label{f1}%
\end{figure}

As shown in Ref. \cite{ccl05}, one can evaluate the fluctuations of the
interferometric phase $\Delta\Phi$ caused by an arbitrary perturbation
$\phi\left(  t\right)  $ by
\begin{equation}
\Delta\Phi=\int_{-\infty}^{+\infty}g\left(  t\right)  \frac{d\phi\left(
t\right)  }{dt}dt.\label{ip}%
\end{equation}
Of course, the influence due to the atomic acceleration will be sensed by the
interferometer through the sensitivity function. In general, this type of
time-domain atomic interferometer discussed here is an accelerometer. In the
free evolution of the atom, its phase changes by $\phi_{g}\left(  t\right)
=\frac{\omega_{a}}{c}x=\frac{\omega_{a}}{c}(x_{0}+v_{0}t+\frac{1}{2}gt^{2})$
where $x_{0}$, $v_{0}$ are the initial position and velocity of the atom. Then
using the Eq. (\ref{ip}), we recover the signal of the interferometer%
\begin{equation}
\Delta\Phi_{s}=\frac{\omega_{a}}{c}gT^{2}+O(1)\label{sfg}%
\end{equation}
where $\omega_{a}$ is the atomic energy level difference and $O(1)$ represents
the terms related to $\tau$ or $\frac{1}{\Omega}$, which are tiny values
compared to the leading term. If the relativistic calculation is considered as
in Ref. \cite{dghk08} and the corresponding geodesic is used to get the phase
disturbance $\phi_{g}\left(  t\right)  $, more effects will be presented. Here
the calculation is feasible, because the phase change of the atom induced by
the gravitational field during the free evolution is equivalent to the phase
change of the laser pulse by the fact that the gravitational field changes the
positions of the interaction between the laser pulses and atoms. On the other
hand, in the free evolution, the phase changes actually are $\phi_{g0}\left(
t\right)  =\frac{\omega_{0}}{c}x$ where $\omega_{0}$ is the energy of atomic
ground state and $\phi_{g1}\left(  t\right)  =\frac{\omega_{1}}{c}x$ where
$\omega_{1}$ is the energy of atomic excited state, but after the first beam
splitter, the state of atom is a superposition of the ground and excited
states, so we only need to consider the change of the relative phase,
$\phi_{g}\left(  t\right)  =\frac{\omega_{a}}{c}x$ where $\omega_{a}%
=\omega_{1}-\omega_{0}$.

\section{Configurational analysis of a new atomic interferometer}

In this section, we will use the sensitivity function to analyze the
configuration suggested in Ref. \cite{ghkr13}, which can be understood as a
variant of a light-pulse de Broglie wave interferometer in Mach-Zender
configuration. This kind of atomic interferometer is derived from the
time-domain atomic interferometer introduced in the last section, but with an
obvious difference by adding some pulses between the basic beam splitters and
the mirror pulse which leads to a large momentum transfer (LMT). All added
pulses are $\pi$ pulses (to distinguish the basic mirror $\pi$ pulse, we call
these specific $\pi$ pulses) which are refined to interact only with the
required halves of atoms, unlike the three basic pulses that interact with all
atoms simultaneously but with different influences on each half. In order to
realize the LMT, these specific $\pi$ pulses must be arranged carefully. Here
we will show that for the given number of the specific $\pi$ pulses, the
momentum transfer will be dependent on \textit{the sequence} (that includes
the consideration of the direction of pulses and which halves of atoms will
interact with the laser pulses) and \textit{the time} (that means what time
each pulse is applied). The case of $N=3$ is used to show this. The sign $N$
is slightly different from that used in Ref. \cite{ghkr13}, and here our signs
are related to the final leading order phase shift for the case with the
largest momentum transfer, e.g. for the interferometer of $N=3$ presented in
Fig.2, it is $\sim3\omega_{a}gT^{2}/c$. In particular, the case presented in
the FIG. 2 of Ref. \cite{ghkr13} is $N=2$ according to our rule of
signs.\bigskip\begin{figure}[ptb]
\centering
\includegraphics[width=3.25in]{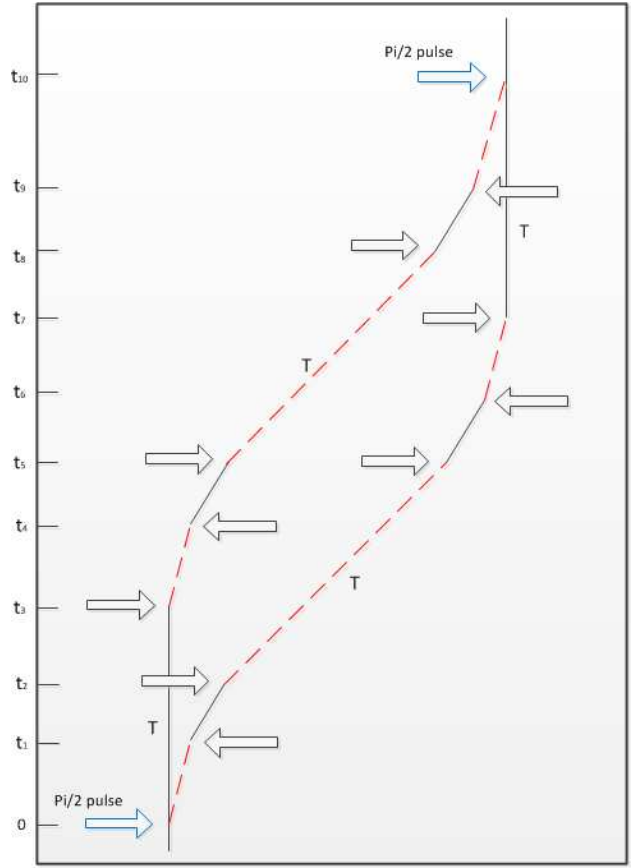}\caption{A spacetime disgram of the
proposed N=3 configuration. The non-labelled arrows represent the $\pi$
pulses, and the directions of the arrows is along the motion of the pulses.
The joint points linking the solid and dashed lines indicate the vertices at
which the laser interacts with the atom. }%
\label{f2}%
\end{figure}

It is seen easily from the interferometer of $N=3$ that there are four
specific $\pi$ pulses before the basic mirror pulse and the arrangement
presented in Fig.2 is the case with the largest momentum transfer. Actually,
there are many other ways to arrange the four $\pi$ pulses, and the final
leading order phase shifts will include the possible results $\sim$
$2\omega_{a}gT^{2}/c$, $\omega_{a}gT^{2}/c$ and $0$. There are $30$ possible
structures in our consideration with different sequences and different times,
but there is only one that can realize the largest momentum transfer, i.e. the
case of the final leading order phase shift is $3\omega_{a}gT^{2}/c$. It is
pointed out that the counting of all possible configurations of
interferometers is based on the assumption that the laser emitter is fixed on
a position or the basic beam splitters are from the same direction. Then there
is a natural problem: how can we know which is LMT and how many momenta are
transferred. One standard method is to calculate all of the interactions
included in the process of the interferometer to get the final leading phase
shift, e.g. using the method in Ref. \cite{dghk08}, but the relativistic
calculation is not required. Here we will provide another interesting method
with the sensitivity function. For the case of $N=3$ in Fig.2, due to the odd
symmetry of the sensitivity function, we write it only for $t<0$,%
\[
g_{3}(t)=\left\{
\begin{array}
[c]{c}%
\sin\Omega(t+T)\ \ \ \ \ \ \ \ \ \ \ \ \ \ ,-T\leqslant t<-T+\tau\\
\ \ \ \ \ \ \ \ \ 1\ \ \ \ \ \ \ \ \ \ \ \ \ \ \ \ \ \ \ ,-T+\tau\leqslant
t<-T+3\tau\\
\frac{1}{2}[3-\cos\Omega(T+t-3\tau)]\ \ \ \ \ \ ,-T+3\tau\leqslant
t<-T+5\tau\\
\ \ \ \ \ \ \ \ \ \ 2\ \ \ \ \ \ \ \ \ \ \ \ \ \ \ \ \ \ \ ,-T+5\tau\leqslant
t<-T+7\tau\\
\frac{1}{2}[5-\cos\Omega(T+t-7\tau)]\ \ \ \ \ \ ,-T+7\tau\leqslant
t<-T+9\tau\\
\ \ \ \ \ \ 3\ \ \ \ \ \ \ \ \ \ \ \ \ \ \ \ \ \ \ ,-T+9\tau\leqslant
t<-9\tau\\
\ \frac{1}{2}[5+\cos\Omega(t+9\tau)]\ \ \ \ \ \ \ \ \ \ ,-9\tau\leqslant
t<-7\tau\ \ \ \ \ \ \ \ \ \\
\ \ 2\ \ \ \ \ \ \ \ \ \ \ \ \ \ \ \ \ \ \ ,-7\tau\leqslant t<-5\tau\\
\ \ \frac{1}{2}[3+\cos\Omega(t+5\tau)]\ \ \ \ \ \ \ \ \ \ ,-5\tau\leqslant
t<-3\tau\ \ \ \ \ \ \ \ \ \ \\
\ 1\ \ \ \ \ \ \ \ \ \ \ \ \ \ \ \ \ \ \ ,-3\tau\leqslant t<-\tau\\
\ -\sin\Omega t\ \ \ \ \ \ \ \ \ \ \ \ \ \ \ \ ,-\tau\leqslant t<0\ \ \ \ \ \
\end{array}
\right.
\]
which is calculated detailedly in the appendix2. Then for the information from
the gravitational field, the leading order leads to,
\begin{equation}
\Delta\Phi_{3}=\int_{-\infty}^{+\infty}g_{3}\left(  t\right)  \frac{d\phi
_{g}\left(  t\right)  }{dt}dt=3\frac{\omega_{a}}{c}gT^{2},\label{3pc}%
\end{equation}
and as expected, it includes the largest momentum transfer. This result is
also seen implicitly from the sensitivity function itself, e.g. $g_{3}\left(
t\right)  =3$, for $-T+9\tau\leqslant t<-9\tau$, as presented in Fig.3.
However, the time that the specific $\pi$ pulses are applied works definitely,
e.g. the time for the eleven pulses calculated here is t1: $\left(  -T\right)
\rightarrow\left(  -T+3\tau\right)  \rightarrow\left(  -T+7\tau\right)
\rightarrow\left(  -9\tau\right)  \rightarrow\left(  -5\tau\right)
\rightarrow\left(  -\tau\right)  \rightarrow\left(  3\tau\right)
\rightarrow\left(  7\tau\right)  \rightarrow\left(  T-9\tau\right)
\rightarrow\left(  T-5\tau\right)  \rightarrow\left(  T-\tau\right)  $. If we
change the times but don't change the sequence, e.g. when the time take t2:
$\left(  -T\right)  \rightarrow\left(  -T+3\tau\right)  \rightarrow\left(
-13\tau\right)  \rightarrow\left(  -9\tau\right)  \rightarrow\left(
-5\tau\right)  \rightarrow\left(  -\tau\right)  \rightarrow\left(
3\tau\right)  \rightarrow\left(  7\tau\right)  \rightarrow\left(
9\tau\right)  \rightarrow\left(  T-5\tau\right)  \rightarrow\left(
T-\tau\right)  $, we have $\Delta\Phi_{3}=2\frac{\omega_{a}}{c}gT^{2}$ with
the same integral as in (\ref{3pc}) but with the different times. Of course,
we can also find the arrangement of the time for the sequence to make
$\Delta\Phi_{3}=\frac{\omega_{a}}{c}gT^{2}$, but $\Delta\Phi_{3}=0$ doesn't
exist for the sequence. Thus given the sequence as in Fig.2, there are $5$
different arrangements of the times.

\bigskip\bigskip\begin{figure}[ptb]
\centering
\includegraphics[width=5.25in]{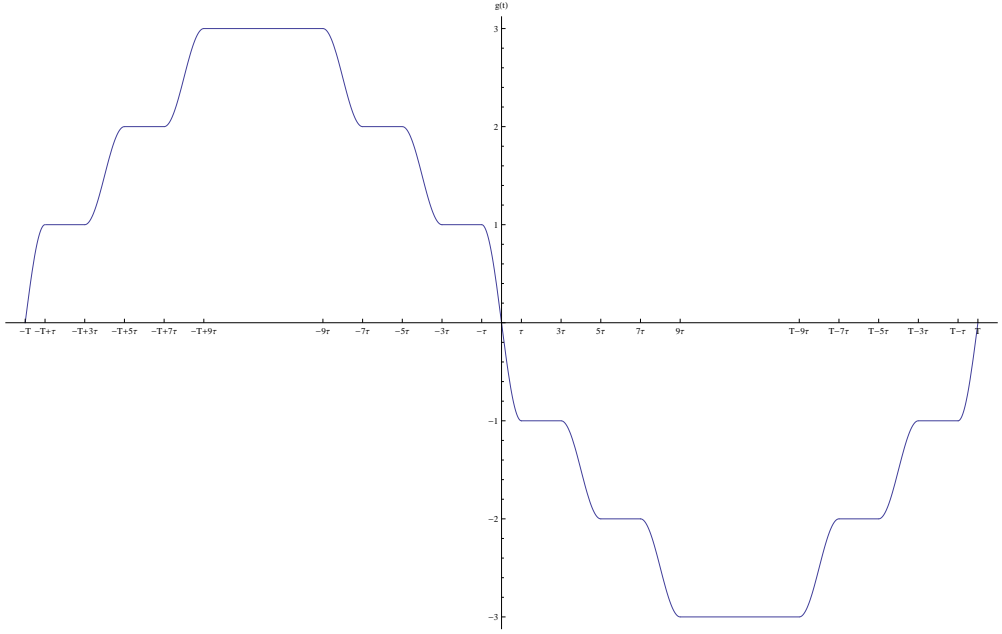}\caption{The sensitivity function
$g_{3}(t)$ for the case of N=3 with the largest momentum transfer. }%
\label{f3}%
\end{figure}

\bigskip

For a given time that the specific $\pi$ pulses are applied, there are six
different sequences in all for the case of $N=3$. The schematics of the
sensitivity functions of the other five sequences are presented in Fig.4. All
the $30$ kinds of cases are summarized in the table 1. It is noted that there
exist some interesting cases, e.g., $\Delta\Phi_{3}=0$ and this is impossible
for the original interferometer consisted of beam splitter-mirror-beam
splitter ($\frac{\pi}{2}-\pi-\frac{\pi}{2}$) optical pulse sequence. More
interesting, for the sequence s6 and the time t1, $\Delta\Phi_{3}%
=-\frac{\omega_{a}}{c}gT^{2}$ which includes a reversed momentum recoil.

\begin{table}[h]
\centering
\begin{tabular}
[c]{|c|c|c|c|c|c|c|}\hline
time/sequence & s1 & s2 & s3 & s4 & s5 & s6\\\hline
t1 & $3\omega_{a}gT^{2}/c$ & $\omega_{a}gT^{2}/c$ & $\omega_{a}gT^{2}/c$ &
$\omega_{a}gT^{2}/c$ & $\omega_{a}gT^{2}/c$ & $-\omega_{a}gT^{2}/c$\\\hline
t2 & $2\omega_{a}gT^{2}/c$ & $2\omega_{a}gT^{2}/c$ & $0$ & $2\omega_{a}%
gT^{2}/c$ & $0$ & $0$\\\hline
t3 & $2\omega_{a}gT^{2}/c$ & $2\omega_{a}gT^{2}/c$ & $2\omega_{a}gT^{2}/c$ &
$0$ & $0$ & $0$\\\hline
t4 & $\omega_{a}gT^{2}/c$ & $\omega_{a}gT^{2}/c$ & $\omega_{a}gT^{2}/c$ &
$\omega_{a}gT^{2}/c$ & $\omega_{a}gT^{2}/c$ & $\omega_{a}gT^{2}/c$\\\hline
t5 & $\omega_{a}gT^{2}/c$ & $\omega_{a}gT^{2}/c$ & $\omega_{a}gT^{2}/c$ &
$\omega_{a}gT^{2}/c$ & $\omega_{a}gT^{2}/c$ & $\omega_{a}gT^{2}/c$\\\hline
\end{tabular}
\caption{We list the results of the final leading phase shift for the
structure of $N=3$. The rows refer to different sequences presented in Fig.4.
The columns refers to different times, i.e. t1 and t2 have been indicated in
the paper; t3:$\left(  -T\right)  \rightarrow\left(  -T+3\tau\right)
\rightarrow\left(  -T+7\tau\right)  \rightarrow\left(  -T+11\tau\right)
\rightarrow\left(  -5\tau\right)  \rightarrow\left(  -\tau\right)
\rightarrow\left(  3\tau\right)  \rightarrow\left(  T-13\tau\right)
\rightarrow\left(  T-9\tau\right)  \rightarrow\left(  T-5\tau\right)
\rightarrow\left(  T-\tau\right)  $; t4:$\left(  -T\right)  \rightarrow\left(
-17\tau\right)  \rightarrow\left(  -13\tau\right)  \rightarrow\left(
-9\tau\right)  \rightarrow\left(  -5\tau\right)  \rightarrow\left(
-\tau\right)  \rightarrow\left(  3\tau\right)  \rightarrow\left(
7\tau\right)  \rightarrow\left(  11\tau\right)  \rightarrow\left(
15\tau\right)  \rightarrow\left(  T-\tau\right)  $; t5:$\left(  -T\right)
\rightarrow\left(  -T+3\tau\right)  \rightarrow\left(  -T+7\tau\right)
\rightarrow\left(  -T+11\tau\right)  \rightarrow\left(  -T+15\tau\right)
\rightarrow\left(  -\tau\right)  \rightarrow\left(  T-17\tau\right)
\rightarrow\left(  T-13\tau\right)  \rightarrow\left(  T-9\tau\right)
\rightarrow\left(  T-5\tau\right)  \rightarrow\left(  T-\tau\right)  $. }%
\label{Table1}%
\end{table}

\begin{figure}[ptb]
\centering
\includegraphics[width=2.5in]{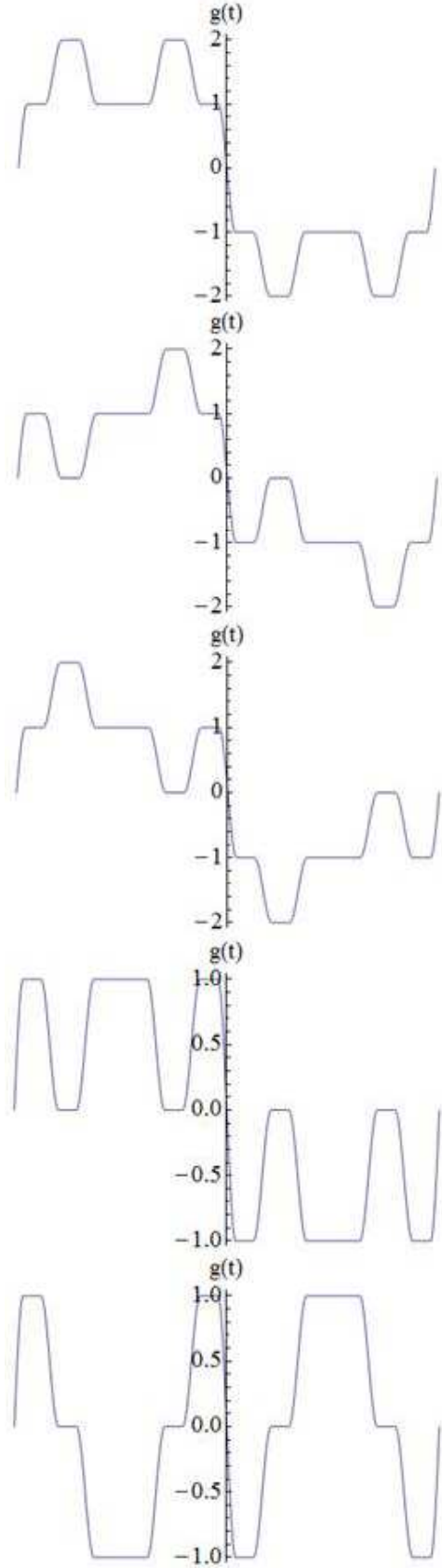}\caption{The schematics of the
sensitivity function for the interferometers with the other five sequences,
respectively. }%
\label{f4}%
\end{figure}

When the $N$ increases, the kinds of possible interferometers will increase.
For example, when $N=4$, there are $140$ kinds with $20$ different sequences
and $7$ different arrangements of the times for each sequence. Notably, for
every kind of interferometer, there is only one structure to realize the
largest momentum transfer, i.e. the final leading order phase shift is
$N\omega_{a}gT^{2}/c$.

It has to be pointed out that when the LMT interferometers amplify the signal,
the noises are also amplified simultaneously, which attenuates the advantage
of such interferometers. This can be seen easily from the white phase noise,
and here we present a result from the vibration noises. Fig.5 is a vibration
spectrum measured in our lab. Assume such vibration happens in the measurement
process of an interferometer, and we could estimate its influence on the final
population change of atoms using the Eq. (\ref{stf}) of the appendix1, i.e.
for the interferometer using three pulses described in the last section, the
estimation of its influence is $\sim2.99\times10^{-7}$; for the interferometer
of $N=3$ described in this section, the estimation is $\sim1.18\times10^{-6}$.
It is seen easily that the influence of the vibration noise is amplified to
nearly 4 times the previous one.

\begin{figure}[ptb]
\centering
\includegraphics[width=3.25in]{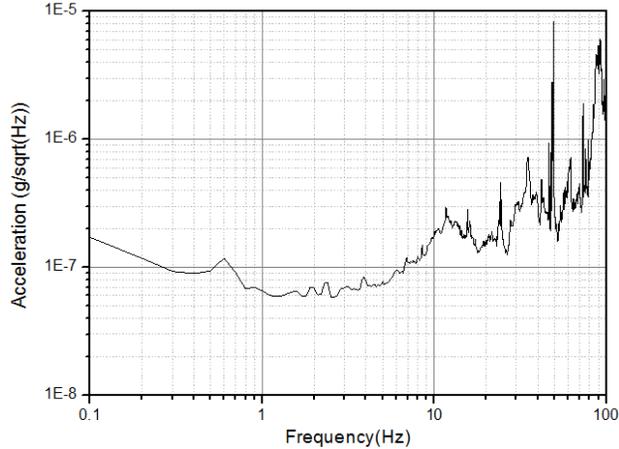}\caption{The spectrum of the vibration
measured in our lab. }%
\label{f5}%
\end{figure}

\section{Configuration for the detection of gravitational waves}

Although the new interferometers with LMT amplify noises existed in the
process of the interferometry, many of the noises will be cancelled when the
two new interferometers are operated in a proper method, as in the case of the
detection of gravitational waves \cite{yt10}. Compared with the above
discussion, the influence of vibration noise is $\sim2.9\times10^{-11}$ for
the configuration with the baseline $L=1000$ km, which presents a very large
suppression. See the schematic in Fig.6, and the consideration is similar to
the Refs. \cite{yt10,ghkr13}, with the atomic interferometers replaced by the
new interferometers discussed in the last section. The laser emitter is laid
at the left side and the secondary pulse is formed by the reflection from the
mirror fixed at the right side. In particular, the time interval between the
primary and secondary pulses depends on the distance between the mirror and
the right interferometer.

\begin{figure}[ptb]
\centering
\includegraphics[width=3.25in]{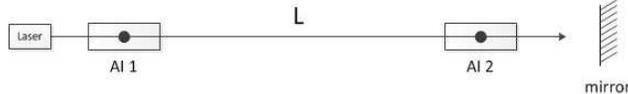}\caption{The schematic of the detector
of gravitational waves with the use of the new interferometers.}%
\label{f6}%
\end{figure}

The related noise analysis had been made in Refs. \cite{dgr08,ghkr13}. In
particular, Ref. \cite{ghkr13} pointed out a remarkable point that the
configuration is immune to laser frequency noise for the detection scheme of
gravitational waves with the use of the new interferometers. Here we find some
subtle differences, which will be discussed below, e.g. the suppression of the
laser frequency noise is different between the cases of $N$ being even and
odd, for the case with the largest momentum transfer (in the section we will
only refer to the case with the largest momentum transfer, except a specific
one is pointed out explicitly).

Firstly, we discuss the case for which $N$ is even, and assume that $N=2$
without loss of generality. This case had been introduced in the FIG.2 of Ref.
\cite{ghkr13} with a $2\hbar k$ momentum transfer, in which the phase change
due to the interaction of the laser with the atom can be expressed as
\[
\Phi_{L}=\phi_{1}+\phi_{2}-\phi_{3}-2\phi_{4}-\phi_{5}+\phi_{6}+\phi_{7},
\]
where the subscripts are associated with the seven pulses in chronological
sequence. Now we consider the phase change caused by the phase fluctuation
$\phi\left(  t\right)  $ which is due to the laser instability (here we ignore
the other noise sources such as the fluctuation from atomic coherence), where
$t$ means the time that the pulse is emitted from the laser emitter since the
phase of a laser does not evolve during its propagation in vacuum. A direct
intuition is to use the TDI to cancel the laser frequency noise, which is also
mentioned in Ref. \cite{ghkr13}, but we find this is not feasible for the
present scheme. Similar to the Ref. \cite{yt10}, the responses $R\left(
t\right)  $ of the new atomic interferometers to the laser phase noise can be
written as%
\begin{align}
R_{21}\left(  t\right)   &  =\phi\left(  t\right)  +\phi\left(  t-3\tau
-L\right)  -\phi\left(  t-T+5\tau\right)  -2\phi\left(  t-T+\tau-L\right)
\nonumber\\
&  -\phi\left(  t-T-3\tau\right)  +\phi\left(  t-2T+5\tau-L\right)
+\phi\left(  t-2T+\tau\right)  ;\label{rfa1}\\
R_{22}\left(  t\right)   &  =\phi\left(  t-L\right)  +\phi\left(
t-3\tau\right)  -\phi\left(  t-T+5\tau-L\right)  -2\phi\left(  t-T+\tau
\right)  \nonumber\\
&  -\phi\left(  t-T-3\tau-L\right)  +\phi\left(  t-2T+5\tau\right)
+\phi\left(  t-2T+\tau-L\right)  ,\label{rfa2}%
\end{align}
where for brevity we take $c=1$, $t$ is chosen at the time that the final
laser is emitted, and \textquotedblleft$2$\textquotedblright\ in the
subscripts of $R_{21}\left(  t\right)  $ represents the case of $N=2$ and
\textquotedblleft$1$\textquotedblright\ represents the left interferometer in
Fig.6. In particular, the subscripts of $\phi$ are omitted, because the laser
frequency noise is only caused by the laser emitter and thus derives from the
same function $\phi\left(  t\right)  $. If the laser is emitted from the left,
the time of interaction between the laser and atoms in the right
interferometer is delayed by $\Delta t=L$, and vice versa. By using the method
of TDI, it was expected that $\Delta R\left(  t\right)  =R_{21}\left(
t-L\right)  -R_{22}\left(  t\right)  =0$, but actually it is not true for the
case we are discussing, because a calculation gives $\Delta R\left(  t\right)
\neq0$ by using the expressions (\ref{rfa1}) and (\ref{rfa2}). This is because
for the present scheme, not all the laser pulses are emitted from the same
direction, which is different from the situation discussed in Ref.
\cite{yt10}. However, we find that, when $\tau$ is small enough compared with
$L$ and $T$ (e.g. in a proposed experiment of Ref. \cite{ghkr13}, $\tau=20\mu
$s, $L=1000$km or $3$ms, $T=1.5$s), the response functions can be written as%
\begin{align}
R_{21}\left(  t\right)   &  =\phi\left(  t\right)  +\phi\left(  t-L\right)
-\phi\left(  t-T\right)  -2\phi\left(  t-T-L\right)  \nonumber\\
&  -\phi\left(  t-T\right)  +\phi\left(  t-2T-L\right)  +\phi\left(
t-2T\right)  ;\\
R_{22}\left(  t\right)   &  =\phi\left(  t-L\right)  +\phi\left(  t\right)
-\phi\left(  t-T-L\right)  -2\phi\left(  t-T\right)  \nonumber\\
&  -\phi\left(  t-T-L\right)  +\phi\left(  t-2T\right)  +\phi\left(
t-2T-L\right)  .
\end{align}
It is found directly that $\delta R(t)=R_{21}\left(  t\right)  -R_{22}\left(
t\right)  =0$. So one doesn't need to consider any time delay, since the laser
frequency noise can be cancelled at any time. Note that the cancellation is
subtle, e.g. the first term of $R_{21}\left(  t\right)  $ is canceled by the
second term of $R_{22}\left(  t\right)  $. This shows that the cancellation is
better when the time internal is shorter (it is best for no time internal as
the presentation in Ref. \cite{ghkr13} where the primary laser is triggered at
time $t=0$, and the secondary one at time $t=L/c$ without any time internal).
This result can also be extended to any case for which $N$ is even, since in
that case, the number of laser pulses from one direction is equal to that from
the other one.

However, we find the case for which $N$ is odd, is not the same. We show this
by using the case of $N=3$ presented in Fig.2, and its responses $R\left(
t\right)  $ of the new atomic interferometers to the laser phase noise can be
expressed as%
\begin{align}
R_{31}\left(  t\right)   &  =\phi\left(  t\right)  +\phi\left(  t-3\tau
-L\right)  +\phi\left(  t-7\tau\right)  -\phi\left(  t-T+9\tau\right)
-\phi\left(  t-T+5\tau-L\right) \nonumber\\
&  -2\phi\left(  t-T+\tau\right)  -\phi\left(  t-T-3\tau-L\right)
-\phi\left(  t-T-7\tau\right)  +\phi\left(  t-2T+9\tau\right) \nonumber\\
&  +\phi\left(  t-2T+5\tau-L\right)  +\phi\left(  t-2T+\tau\right)
\label{rfln1}\\
R_{32}\left(  t\right)   &  =\phi\left(  t-L\right)  +\phi\left(
t-3\tau\right)  +\phi\left(  t-7\tau-L\right)  -\phi\left(  t-T+9\tau
-L\right)  -\phi\left(  t-T+5\tau\right) \nonumber\\
&  -2\phi\left(  t-T+\tau-L\right)  -\phi\left(  t-T-3\tau\right)
-\phi\left(  t-T-7\tau-L\right)  +\phi\left(  t-2T+9\tau-L\right) \nonumber\\
&  +\phi\left(  t-2T+5\tau\right)  +\phi\left(  t-2T+\tau-L\right)  .
\label{rfln2}%
\end{align}
Again it is found that $\Delta R\left(  t\right)  =R_{31}\left(  t-L\right)
-R_{32}\left(  t\right)  \neq0$ even though small $\tau$ is considered. In
particular, it is also found that $\delta R(t)=R_{31}\left(  t\right)
-R_{32}\left(  t\right)  \neq0$ when the time internal $\tau$ is ignored; this
is seen as%
\[
\delta R(t)=\phi\left(  t\right)  -\phi\left(  t-L\right)  -2\left(
\phi\left(  t-T\right)  -\phi\left(  t-T-L\right)  \right)  +\phi\left(
t-2T\right)  -\phi\left(  t-2T-L\right)  \text{.}%
\]
As expected, $\delta R(t)$ is very small since $\phi\left(  t\right)  $ is a
slowly varying function which is required for any laser emitter. On the other
hand, since $T>>L$, we might have $\delta R(t)=0$ by omitting the time
internal $L$. That is permitted if it is within the sensitivity of the
interferometer. But for the detection of gravitational waves, it is expected
to remain the influence happened in the time internal $L$ since the signal is
proportional to the length $L$. It is found that the reason that laser
frequency noise cannot be cancelled exactly for any case with $N$ being odd is
due to the asymmetry of the number of laser pulses from the two directions.
However, the laser frequency noise is suppressed to a large extent, in the
scheme with LMT.

Finally, we will point out the difference of $\Delta R\left(  t\right)  $ and
$\delta R(t)$ operationally. $\Delta R\left(  t\right)  $ means that for each
interaction, the pulse is the same for the first and the second
interferometer. Such operation is easy to be understood physically for the
measurement of gravitational waves. In particular, it also gets a large
suppression of laser frequency noise in the present scheme. The reason that we
use $\delta R(t)$ is that it leads to a complete cancellation of the laser
frequency noise for the cases of even $N$. But for a single interaction, such
operation means the pulses used to interact with the atoms are different for
the first and second interferometers. Of course, it is only a choice of a
proper data processing method \cite{dam}, but we have to explain whether such
operation can give the signal of gravitational waves. Fortunately, the latter
operation will not change the signal since the response of the first
interferometer to the gravitational waves, $R_{GW}(t)$, is usually treated as
zero \cite{yt10}. Thus $\delta R_{GW}(t)=R_{1GW}(t)-R_{2GW}(t)=$
$R_{1GW}(t-L)-R_{2GW}(t)=$ $\Delta R_{GW}\left(  t\right)  $.

\section{Conclusion}

In this paper we have investigated the sensitivity function and applied it to
the transfer of the signal about gravitational field around the Earth through
an atomic interferometer. We have also extended the application of sensitivity
function to a new atomic interferometer for which we have analyzed its
configuration in detail. In our analysis, we found that given the number of
laser pulses, there are some different methods to realize LMT although the
transferred momenta in each method are different. In particular, there is only
one method that can realize the largest momentum transfer (i.e. $N\hbar k$
momenta transfer for a scheme with the specific $N$ defined in the paper), and
in all these methods there are some configurations that gives the results of
zero momentum transfer and even inverse momentum transfer (e.g. the final
leading phase shift is $0$ or $-\frac{\omega_{a}}{c}gT^{2}$). For the
configuration in Fig.6 for the detection of gravitational waves, we have also
analyze how the laser frequency noise is cancelled. We found that when $N$ is
even, the configuration is immune to the laser frequency noise with a proper
data processing method; when $N$ is odd, the configuration only gives a large
suppression for the laser frequency noise. However, for the present situation
of detection scheme of gravitational waves, the use of new atomic
interferometers with LMT is still a good progress for the amelioration of the
laser frequency noise.

\section{Acknowledgements}

Financial support from NSFC under Grant Nos. 11104324, 11374330 and 11227803,
and NBRPC under Grant Nos.2010CB832805 is gratefully acknowledged.

\section{Appendix1}

In the appendix, we will introduce briefly the transfer function.

In the paper, we used the sensitivity function to analyze the structure of an
interferometer, but actually, maybe a little surprised, it is noted that when
we choose $t_{i}=-T-\tau$ and $t_{f}=T+\tau$, the expression of the
sensitivity function becomes%
\begin{equation}
g_{c}(t)=\left\{
\begin{array}
[c]{c}%
\cos\Omega(T+t)\ \ \ \ ,-T-\tau\leqslant t<-T\\
\ 1\ \ \ \ \ \ \ \ \ \ \ \ \ \ ,-T\ \leqslant t<-\tau\ \ \ \ \\
-\sin\Omega t\ \ \ \ \ \ \ \ \ ,-\tau\leqslant t<\tau\ \ \ \ \ \ \\
-1\ \ \ \ \ \ \ \ \ \ \ \ \ ,\tau\leqslant t<T\ \ \ \ \ \ \ \\
-\cos\Omega(T-t)\ \ \ ,T\leqslant t\leqslant T\ +\tau\ \
\end{array}
\right.  .\label{sf2}%
\end{equation}
Thus if the initial time or the final time is chosen with a change of $\tau$,
during that time the sensitivity function will change from sinusoidal to
cosinusoidal function. This is easy to be known that such change is because we
take $\Omega\tau=$ $\frac{\pi}{2}$. But the transfer function is nearly the
same for the two different choices, which is a mathematical representation of
the relation between the input and output of a measurement system and is also
usually called the weighting function.

For the interferometer we considered here, the Fourier transform of the
sensitivity function is
\[
G\left(  \omega\right)  =\int_{-\infty}^{+\infty}e^{-i\omega t}g\left(
t\right)  dt.
\]
and using sensitivity functions (\ref{sf1}), we have
\begin{equation}
G\left(  \omega\right)  =\frac{4i\Omega^{2}}{\Omega^{2}-\omega^{2}}\sin\left(
\frac{\omega T}{2}\right)  \left[  \frac{1}{\Omega}\cos\left(  \frac{\omega
T}{2}\right)  +\frac{1}{\omega}\sin\left(  \frac{\omega\left(  T-2\tau\right)
}{2}\right)  \right]  .\label{sfft}%
\end{equation}
In particular, for $\tau\ll T$, we have $G_{c}\left(  \omega\right)  \simeq
G\left(  \omega\right)  $. Thus although our paper analyze the structure of
the interferometers using the sensitivity function without discussing its
dependence on the choice of the initial and final time, all results are
applicable to any other analysis related to any change of the initial and
final time.

Similarly, we can express the Fourier transform of the phase perturbation as
$\Phi\left(  \omega\right)  =$ $\int_{-\infty}^{+\infty}e^{-i\omega t}%
\phi\left(  t\right)  dt$ and then put the reverse transform $\phi\left(
t\right)  =\int_{-\infty}^{+\infty}e^{i\omega t}\Phi\left(  \omega\right)
d\omega$ into Eq. (\ref{ip}). Thus we get%
\begin{equation}
\Delta\Phi=-\int_{-\infty}^{+\infty}i\omega G\left(  \omega\right)
\Phi\left(  \omega\right)  d\omega. \label{fip}%
\end{equation}
Since the Fourier form $G\left(  \omega\right)  $ includes only a pure
imaginary part seen from the Eq. (\ref{sfft}), the interferometric phase is
also written as $\Delta\Phi=\int_{-\infty}^{+\infty}\omega\left\vert G\left(
\omega\right)  \right\vert \Phi\left(  \omega\right)  d\omega$. After the
introduction of the transfer function
\begin{equation}
H\left(  \omega\right)  =\omega G\left(  \omega\right)  , \label{tf}%
\end{equation}
we have%
\begin{equation}
\Delta\Phi=\int_{-\infty}^{+\infty}\left\vert H\left(  \omega\right)
\right\vert \Phi\left(  \omega\right)  d\omega, \label{ftip}%
\end{equation}
which presents clearly the process how the interferometer responds to the
phase disturbance (of course it also includes the signal). It is seen that if
we put the Fourier transform of $\phi_{g}\left(  t\right)  $ into the Eq.
(\ref{ftip}), we can obtain the same result as Eq. (\ref{sfg}). A notable
feature of the transfer function $H\left(  \omega\right)  $ is a low pass
first order filtering, having an oscillating behavior with a periodic
frequency of $\delta\omega=\frac{2\pi}{T}$ but under the cutoff frequency
$\omega_{c}=\frac{\sqrt{3}\Omega}{3}$ \cite{ccl05}.

However, the Eq. (\ref{ftip}) is not usually used because most studies focused
on the noise analysis while the power spectra of many noises are easier
described, so the most popular is the variance of the phase fluctuation,%
\begin{equation}
\sigma_{\Phi}^{2}=\int_{0}^{+\infty}\left\vert H\left(  \omega\right)
\right\vert ^{2}S_{\phi}\left(  \omega\right)  d\omega\label{stf}%
\end{equation}
where $S_{\phi}\left(  \omega\right)  $ is the power spectral density of the
phase perturbation and usually expressed as $S_{\phi}\left(  \omega\right)
=\left\vert \Phi\left(  \omega\right)  \right\vert ^{2}$. It is stressed that
$\sigma_{\Phi}^{2}$ is not related directly to the fluctuations of the
interferometric phase, while related to the average value of $\Delta\Phi$. In
particular, the average is usually taken for a long sequence of measurement
cycles at a fixed repetition rate which will lead to an aliasing phenomenon
similar to the Dick effect \cite{gjd87} in atomic clocks and increase the
sensitivity of the interferometer on the low frequencies.

An immediate examination is to use the white phase noise with no frequency
dependence, $S_{\phi}\left(  \omega\right)  =S_{\phi}^{0}$, and the result was
given in the Ref. \cite{ccl05} and had a linear dependence on the inverse
Raman pulse length. But the pulse length will affect the number of the
participating atoms which leads to a shot noise and limits the sensitivity of
the interferometer, so the optimum pulse length must be selected with the
reference to the experimental parameters.

\section{Appendix2}

In the appendix, we will give a detailed calculation for the sensitivity
function of the interferometer presented in Fig.2.

Considering the atom which will go through the interferometer as a two-level
system whose general state can be expressed as $\left\vert \varphi\left(
t\right)  \right\rangle =c_{a}\left(  t\right)  \left\vert a\right\rangle
+c_{b}\left(  t\right)  \left\vert b\right\rangle $ with $\left\vert
c_{a}\left(  t\right)  \right\vert ^{2}+\left\vert c_{b}\left(  t\right)
\right\vert ^{2}=1$, the evolution of the state under the interaction with the
laser pulse is calculated with the change of the coefficients \cite{tr05},%
\begin{align}
c_{a}\left(  t_{0}+\tau\right)   &  =c_{a}\left(  t_{0}\right)  \cos\left(
\frac{\Omega\tau}{2}\right)  -ic_{b}\left(  t_{0}\right)  e^{i\phi}\sin\left(
\frac{\Omega\tau}{2}\right)  ,\nonumber\\
c_{b}\left(  t_{0}+\tau\right)   &  =c_{b}\left(  t_{0}\right)  \cos\left(
\frac{\Omega\tau}{2}\right)  -ic_{a}\left(  t_{0}\right)  e^{-i\phi}%
\sin\left(  \frac{\Omega\tau}{2}\right)  ,\label{ce}%
\end{align}
where $\Omega$ is the Rabi frequency, $\tau$ is the duration of the
interaction, and $\phi$ is the relative phase change during the interaction.

For the case of $N=3$, there are eleven pulses, and in order to calculate the
sensitivity function, except the phase $\phi_{0}$ carried by the pulses
themselves which could be tuned in an experiment, a random phase change
$\delta\phi$ during each interaction between the pulse and the atom may be
introduced. For an illustration, we will calculate the result caused by the
phase change $\delta\phi$ at time $t$ during the first pulse, $-T<t<-T+\tau$,
by splitting the pulse into two pulses of duration $T+t$ and $-T+\tau-t$. We
start by assuming all atoms in the state $\left\vert a\right\rangle $, and
according to the sequence presented in Fig.2, the change of the coefficients
after each pulse becomes,%
\[
c_{a1}(-T)=1,
\]%
\[
c_{b2}(-T)=0;
\]%
\[
c_{a1}(t)=\cos\frac{\Omega(t+T)}{2},
\]%
\[
c_{b2}(t)=-ie^{-i\phi_{01}}\sin\frac{\Omega(t+T)}{2};
\]%
\begin{align*}
c_{a1}(-T+\tau) &  =c_{a1}(t-t-T+\tau)\\
&  =c_{a1}(t)\cos\frac{\Omega(\tau-t-T)}{2}-e^{i(\delta\phi+\phi_{01})}%
c_{b2}(t)\sin\frac{\Omega(\tau-T-t)}{2}\\
&  =\frac{\sqrt{2}}{2}(\cos^{2}\alpha+\sin\alpha\cos\alpha)-\frac{\sqrt{2}}%
{2}e^{i\delta\phi}(\sin\alpha\cos\alpha-\sin^{2}\alpha),
\end{align*}%
\[
c_{b2}(-T+\tau)=-\frac{\sqrt{2}}{2}ie^{-i\phi_{01}}(\sin^{2}\alpha+\sin
\alpha\cos\alpha)-\frac{\sqrt{2}}{2}ie^{-i(\phi_{01}+\delta\phi)}(\cos
^{2}\alpha-\sin\alpha\cos\alpha),
\]
where $\alpha=\frac{\Omega(t+T)}{2}$, $\Omega\tau=\frac{\pi}{2}$ for the first
pulse with the change by a random phase $\delta\phi$, and the subscripts $a$,
$b$ represent the two energy levels and 1, 2 represent the different paths.
Since there is no new random phase introduced in the following ten pulses, the
change of the coefficients is calculated easily according to the evolution
(\ref{ce}). For the final pulse, we obtain the expression of $c_{a}$ as%
\begin{align*}
c_{a}(T) &  =\frac{\sqrt{2}}{2}c_{a2}(T-3\tau)-\frac{\sqrt{2}}{2}%
ie^{i(\phi_{11}+\delta\phi)}c_{b1}(T-3\tau)\\
&  =-\frac{\sqrt{2}}{2}ie^{i(-\phi_{02}-\phi_{03}+\phi_{06}+\phi_{07}%
+\phi_{08}+\delta\phi)}c_{b}(-T+\tau)\\
&  -\frac{\sqrt{2}}{2}e^{i(\phi_{09}-\phi_{04}-\phi_{05}-\phi_{06}+\phi
_{10}+\phi_{11})}c_{a}(-T+\tau)\\
&  =-\frac{1}{2}e^{i(-i\phi_{01}-\phi_{02}-\phi_{03}+\phi_{06}+\phi_{07}%
+\phi_{08}+\delta\phi)}(\sin^{2}\alpha+\sin\alpha\cos\alpha)\\
&  -\frac{1}{2}e^{i(-i\phi_{01}-\phi_{02}-\phi_{03}+\phi_{06}+\phi_{07}%
+\phi_{08})}(\cos^{2}\alpha-\sin\alpha\cos\alpha)\\
&  -\frac{1}{2}e^{i(\phi_{09}-\phi_{04}-\phi_{05}-\phi_{06}+\phi_{10}%
+\phi_{11})}(\cos^{2}\alpha+\sin\alpha\cos\alpha)\\
&  +\frac{1}{2}e^{i(\phi_{09}-\phi_{04}-\phi_{05}-\phi_{06}+\phi_{10}%
+\phi_{11}+\delta\phi)}(\sin\alpha\cos\alpha-\sin^{2}\alpha).
\end{align*}
Define $\Phi=\phi_{01}+\phi_{02}+\phi_{03}-\phi_{04}-\phi_{05}-2\phi_{06}%
-\phi_{07}-\phi_{08}+\phi_{09}+\phi_{10}+\phi_{11}=\frac{\pi}{2}$, and we get
the final probability of finding the atom still in state $\left\vert
a\right\rangle $,%
\begin{align*}
P_{a} &  =\left\vert c_{a}(T)\right\vert ^{2}\\
&  =\frac{1}{4}[2+\cos\Phi(1+\cos4\alpha+2\cos\delta\phi\sin^{2}2\alpha
)+2\sin\Phi\sin2\alpha\sin\delta\phi]\\
&  =\frac{1}{2}(1+\sin\Omega(t+T)\sin\delta\phi).
\end{align*}
Thus according to the definition (\ref{sf}) of the sensitivity function, we
have it in the time interval $-T<t<-T+\tau$ as
\[
g(t)=\sin\Omega(t+T).
\]

Then we can proceed in a similar way but the random phase change is introduced
in other ten pulses to obtained the sensitivity functions for those time
intervals, and thus the whole sensitivity function $g_{3}(t)$ is gotten as
presented in the third section of this paper.

The transfer function can be obtained using (\ref{tf}),%
\begin{align*}
H(\omega) &  =\frac{2\omega^{2}i}{\Omega^{2}-\omega^{2}}[\frac{\Omega}{\omega
}\sin\omega T-\cos\omega(T-\tau)+\cos\omega\tau]+2i\sin\omega\tau\lbrack
2\sin\omega(T-2\tau)\\
&  +3\sin\omega(T-4\tau)+4\sin\omega(T-6\tau)+5\sin\omega(T-8\tau
)+5\sin8\omega\tau+4\sin6\omega\tau\\
&  +3\sin4\omega\tau+2\sin2\omega\tau]+12i\sin\frac{\omega(T-18\tau)}{2}%
\sin\frac{\omega T}{2}\\
&  +\frac{2\omega^{2}i}{\Omega^{2}-\omega^{2}}\cos\omega\tau\lbrack-\cos
\omega(T-4\tau)-\cos\omega(T-8\tau)+\cos8\omega\tau+\cos4\omega\tau],
\end{align*}
which is seen from the Fig.7.

\begin{figure}[ptb]
\centering
\includegraphics[width=3.25in]{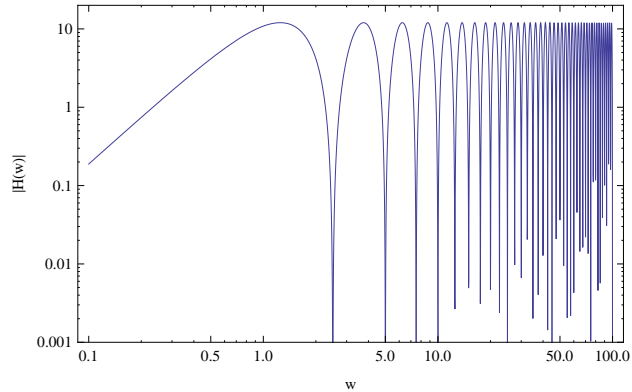}\caption{The transfer function H(w)
which is the Fourier transform of the sensitivty function $g_{3}(t)$ }%
\end{figure}


\begin{thebibliography}{99}
\bibitem {dam}It is stressed that the data processing method is only
applicable for the cancellation of laser frequency noise and whether it is
feasible for the suppression of other noises has to be analyzed independently
and carefully.
\end{thebibliography}
\end{document}